\documentclass[reprint, superscriptaddress, amsmath, amssymb, aps, prx]{revtex4-2}

\usepackage{graphicx}
\usepackage{dcolumn}
\usepackage{bm}
\usepackage{amsmath}	
\usepackage{comment}
\usepackage{makecell}
\usepackage{rotating}   
\usepackage{array}      
\usepackage{tabularx}   

\begin{document}

\preprint{APS/123-QED}

\title{Noise-Resilient Imaging through Coherence Filtering}

\author{Pranay Mohta}
\email{mohtapranay01@gmail.com}
\affiliation{Department of Physics, Indian Institute of Technology Kanpur, Kanpur, UP 208016, India}

\author{Keval Moliya}
\affiliation{Department of Physics, Indian Institute of Technology Kanpur, Kanpur, UP 208016, India}

\author{Aniket Nag}
\affiliation{Department of Physics, Indian Institute of Technology Kanpur, Kanpur, UP 208016, India}

\author{Shaurya Aarav}
\affiliation{Sorbonne Universit\'e, CNRS, Institut des NanoSciences de Paris, INSP, F-75005 Paris, France}

\author{Anand K. Jha}
\email{akjha@iitk.ac.in}
\affiliation{Department of Physics, Indian Institute of Technology Kanpur, Kanpur, UP 208016, India}

\date{\today}

\begin{abstract}
Noise is a significant challenge in imaging. Conventional intensity-based techniques mitigate noise through various filtering methods, but they often require prior knowledge of noise characteristics and struggle, especially under low-light conditions and with spatially structured noise. Quantum distillation provides enhanced noise rejection; however, its applicability is limited as it requires specialised illumination and substantial modifications to existing imaging setups. In this article, we introduce a coherence-based image distillation approach that separates object from noise by leveraging the difference in their temporal coherence properties. We implement this through our interferometric protocol, which enables imaging based on spatial coherence while simultaneously filtering out noise via temporal coherence. This overcomes the limitations of both intensity-based and quantum distillation methods. We experimentally demonstrate noise resilience by successfully recovering feature-rich objects, such as QR codes and grayscale wheels, obscured by spatially uniform and structured noise 20 times as intense as the object. We further show that our method remains effective for fields with substantial spectral overlap, outperforming spectral filtering in regimes where the latter provides little noise suppression. This approach provides a robust framework for noise-resilient imaging with applications in optical communication, fluorescence microscopy, and biological imaging at both high and low light levels.

\end{abstract}

\maketitle


\section{\label{sec:level1}Introduction}

In microscopy and imaging, noise is a ubiquitous challenge that complicates data interpretation, reduces image quality, and hinders accurate quantification. Noise can originate from various sources, such as background illumination, scattering, autofluorescence and sample heterogeneity. To mitigate this, various noise reduction techniques have been developed. Optical filters, like bandpass filters, and the use of pinholes in confocal microscopy, reduce background noise by eliminating unwanted wavelengths and out-of-focus light. However, these methods often lead to signal loss, particularly in low-light or low-contrast conditions, and sacrifice experimental flexibility \cite{pawley2006handbook}. Signal averaging, which reduces random noise by averaging multiple images, works well for stochastic noise but struggles with systematic or structured background noise \cite{gonzalez2009digital}. Post-processing techniques, including background subtraction \cite{donoho2002ieeeit}, Gaussian filtering \cite{jain2013iete,awate2006ieee,shin2005ieee}, median filtering \cite{yang1995ieee,li2009icisp}, wavelet denoising \cite{xu1994ieee,dixit2014,lang1996ieee,silva2012} and others improve the signal-to-noise ratio. However, these methods can distort fine details or fail when the noise is spatially varying or has a similar intensity to the signal \cite{fan2019,goyal2020if,mafi2019sp}. Despite their utility, all of these methods face the challenge of balancing noise reduction with signal preservation, particularly in complex imaging scenarios.

While intensity-based techniques offer some noise mitigation, their limitations have prompted the exploration of quantum imaging techniques. Recent advancements in quantum imaging have introduced protocols designed to provide greater noise resilience. These methods typically involve illuminating the object with entangled photon pairs, while environmental noise is introduced through classical illumination. The images from both sources are then combined on a detector. Early approaches \cite{zhang2020pra,gregory2020scadv,gregory2021scr,zhao2022oe,defienne2019scadv,england2019pra} employed joint detection of photon pairs, leveraging the spatial and temporal correlations between photons to extract the object signal while suppressing noise. While promising, these techniques are inefficient in both the generation and detection of entangled photons, requiring the identification and removal of accidental coincidences, which is both complex and time-consuming. An alternative approach, proposed in Ref.~\cite{fuenzalida2023scadv}, uses single-photon detection for image distillation, using a two-photon effect known as induced coherence. However, this approach too has critical limitations: the optical field of the object passes through the imaging system, whereas the noise field does not. Instead, the noise is mixed with the object field only at the camera. This disparity is a significant challenge because, in real-world scenarios, generally both the object and noise fields pass through the entire imaging system. In quantum distillation techniques, the noise field exhibits different quantum correlations compared to the object field, which is not representative of the typical noise encountered in most imaging and microscopy applications. Furthermore, these techniques are not easily adaptable to standard imaging or microscopy setups and are unsuitable for certain applications, such as fluorescence microscopy. These limitations highlight a key gap in current noise-resilient imaging strategies.
\begin{figure}[t!]
\centering
\includegraphics[scale=0.39]{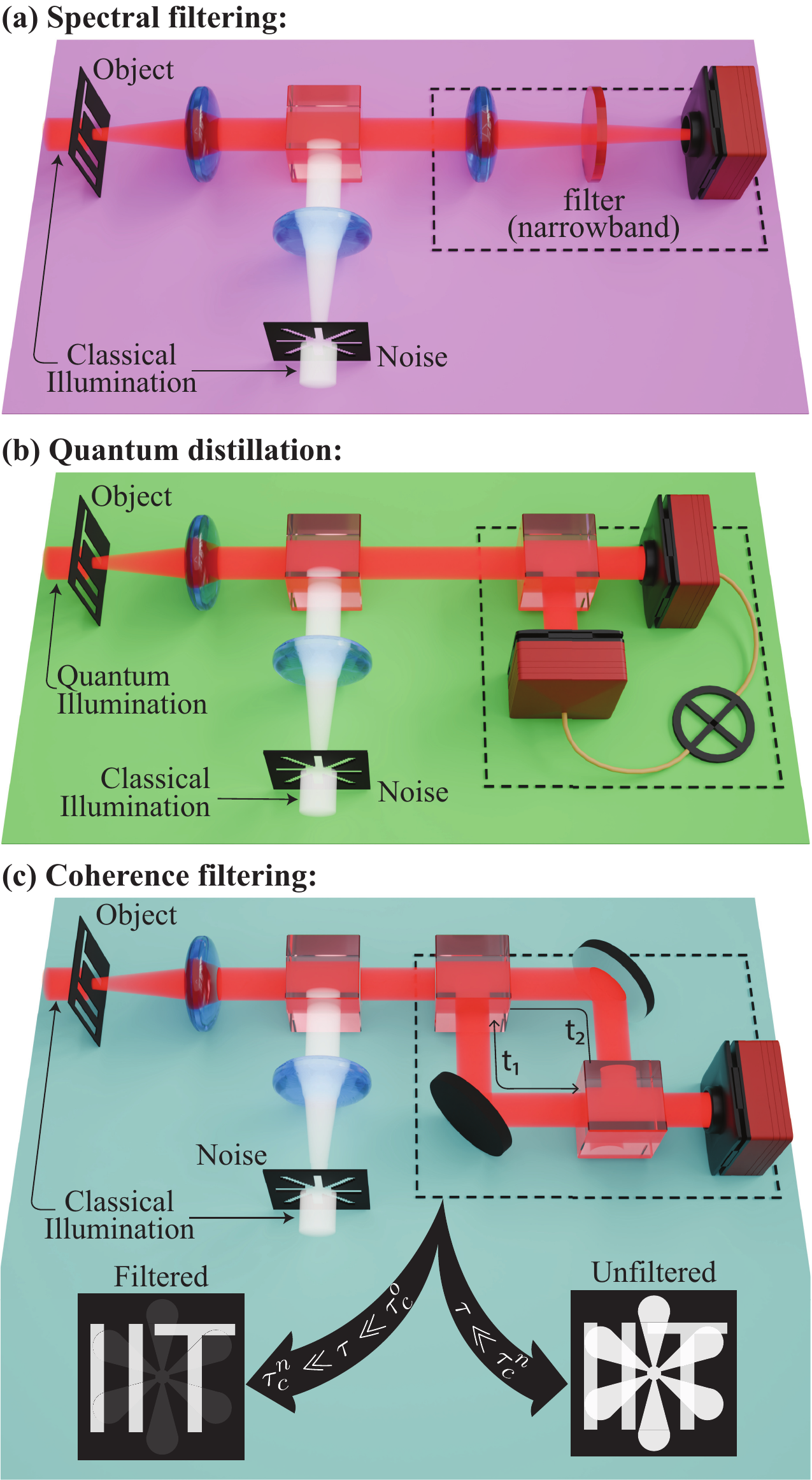}
\caption{Conceptual illustration of noise-resilient imaging techniques - (a) Conventional Intensity imaging with optical filter-based spectral filtering (b) Quantum illumination with coincidence detection (c) Coherence imaging with temporal coherence filtering}
\label{noise-resilient_imaging_schemes_illustration}
\end{figure}
\begin{table}[t]
\caption{Comparison of the filtering techniques}
\centering
\begin{tabular}{|p{1.5cm}|p{2.1cm}|p{4.9cm}|}
\hline
\textbf{Method} & \multicolumn{1}{c|}{\textbf{Principle}} & \multicolumn{1}{c|}{\textbf{Key trade-offs}}\\
\hline
\textbf{Spectral Filtering} &
Frequency discrimination using optical filters. &
\makecell[tl]{Effective when object and noise\\fields are spectrally well separated;\\
Signal loss increases with spectral\\overlap.} \\
\hline
\textbf{Quantum Distillation} &
Temporal and spatial correlations of entangled photon pairs. &
\makecell[tl]{Effective when the object is\\illuminated with entangled photon\\pairs while noise field is classical;\\
High-complexity coincidence\\detection.} \\
\hline
\textbf{Temporal Coherence Filtering} &
Coherence-time discrimination via interferometric path-length delay $\Delta z$. &
\makecell[tl]{Effective when the coherence time\\of the object field is longer than\\that of the noise field;\\
Works for spectrally overlapping\\fields;\\
Requires interferometric stability.} \\
\hline
\end{tabular}
\label{tab:comparison}
\end{table}

We present a filtering technique that separates objects based on the temporal coherence properties of their fields. This filtering technique works in conjunction with coherence imaging, an imaging scheme that measures the spatial coherence of an optical field to construct an image of the object. The idea of coherence imaging~\cite{breckinridge1972ao,itoh1986josaa,yoshimori1997josaa} was initially used in astronomical imaging to overcome the effect of atmospheric turbulence~\cite{breckinridge1972ao,breckinridge1975josaa,roddier1979sn,roddier1985aj,roddier1988pr} and later in microscopy for high depth-of-field imaging~\cite{potuluri2001oe,weigel2015oc}. Typically, light from the object and noise exhibit significantly different coherence characteristics. This disparity is evident in various real-world scenarios. For example, in fluorescence microscopy, the signal originates from specific fluorophores that emit light with a narrow spectral bandwidth around the emission peak, while noise comes from autofluorescence and scattered light, which span a broad spectral range. This difference in spectral bandwidth results in very different temporal coherence properties. Similarly, in free-space imaging/communication, sunlight—characterized by a broad frequency spectrum—serves as the primary source of noise, while the message signal is narrow-band, leading to differing temporal coherence properties.

In this work, we utilize the mismatch in temporal coherence to reliably separate the object signal from noise. Through our interferometric measurement protocol, we demonstrate this imaging approach that leverages spatial coherence for reconstruction while employing temporal coherence for noise rejection, thus establishing a noise-resilient framework for  imaging

\begin{figure*}[t!]
\centering
\includegraphics[scale=0.5]{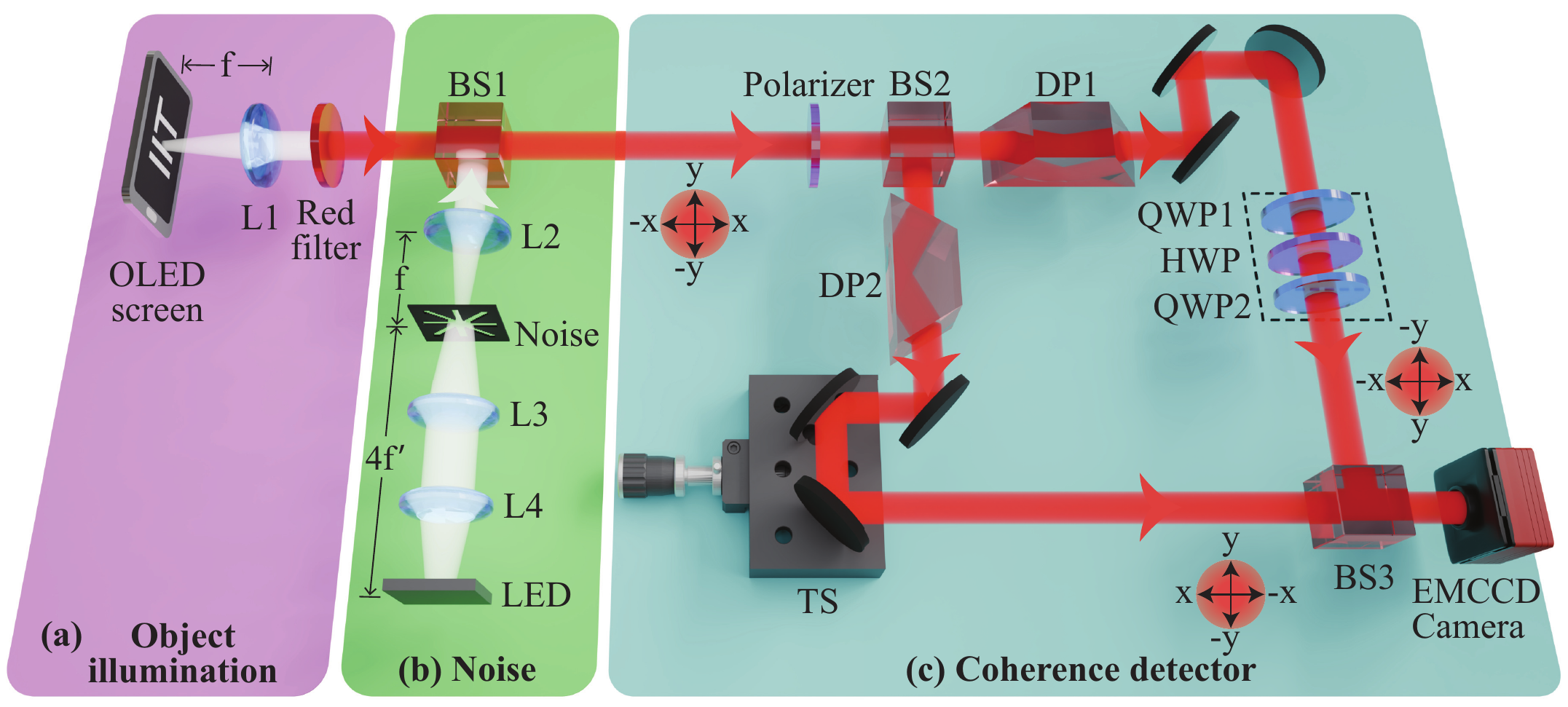}
\caption{Schematic diagram of the experimental setup for Coherence Imaging - 
(a) Object illumination: Object - Image displayed on the OLED screen of a smartphone forms the object illuminated with spatially incoherent light. It is placed at the front focal plane of a converging lens L1 with focal length $f=500\text{mm}$ to get an object field that is spatially and temporally stationary. A red filter centred at 633 nm with a wavelength bandwidth of $10$nm is used to ensure that the object field is narrow-band. (b) Noise - An LED source with a wavelength bandwidth of $90$nm, centred at $595$nm, imaged onto a transparent object forms the noise illuminated with spatially incoherent light. It is placed at the front focal plane of a converging lens L4 with a focal length of $f=500\text{mm}$ to obtain a noise field that is both spatially and temporally stationary. (c) Detector: Phase-shifting wavefront inversion interferometer - It consists of one Dove prism in each arm, at $90^o$ to each other to invert the wavefront of the optical field. One arm contains a geometric phase unit with QWP1, HWP, and QWP2, through which a precise phase difference is introduced between the two arms. The other arm contains a linear translation stage with 0.5 µm precision, used to set the path length difference between the arms. L: thin lenses, BS: beam splitter, DP: dove prism, QWP: quarter-wave plate, HWP: half-wave plate and TS: linear translation stage.}
\label{coherence_imaging_setup}
\end{figure*}
\section{\label{sec:level1}Results}
\subsection{\label{sec:level2}Conceptual illustration of noise-resilient imaging techniques}
Fig.~\ref{noise-resilient_imaging_schemes_illustration} illustrates different noise-resilient imaging schemes. Fig.~\ref{noise-resilient_imaging_schemes_illustration}(a) shows intensity imaging with an optical bandpass filter at the camera to reduce noise. When the spectral bandwidth of the object field is narrower than that of the noise field, the bandpass filter helps reduce noise by blocking unwanted wavelengths. This approach is most effective when the object and noise fields are spectrally well separated; as their spectra overlap, increasing portions of the object field fall within the rejected band, leading to greater signal loss. Fig.~\ref{noise-resilient_imaging_schemes_illustration}(b) illustrates the quantum distillation technique. In this approach, the object is illuminated with entangled photons, while the noise field is classical. Unlike the uncorrelated photons emitted by the classical source, photon pairs from SPDC illumination are temporally and spatially correlated. These correlations are leveraged to extract the image of the object from a mixed quantum-classical image. The noise rejection in this scheme is independent of the spectral overlap between the object and noise fields. Fig.~\ref{noise-resilient_imaging_schemes_illustration}(c) shows our temporal coherence filtering scheme that introduces a coherence-based image distillation approach, achieving noise resilience without requiring any prior information about the object or noise characteristics or the need for specialised quantum illumination.
For interference to occur, the optical field must exhibit both temporal and spatial coherence. The temporal coherence length is inversely related to the spectral width of the field. As illustrated in  Fig.~\ref{noise-resilient_imaging_schemes_illustration}(c), when the spectral bandwidth of the object field is narrower than that of the noise field, the object field has a longer temporal coherence length. If the time difference between the interferometer arms, $\tau = t_2 - t_1$, is much smaller than the noise coherence time ($\tau^n_c$), a high-visibility interference pattern is produced for both the object and noise fields, resulting in a noisy unfiltered image. However, if the time difference is much larger than the noise coherence time ($\tau^n_c$) but still smaller than the object coherence time ($\tau^o_c$), the interference pattern will appear only for the object field, producing a filtered image of the object. This filtering scheme is independent of the spectral overlap between the fields and can reject noise as long as the temporal-coherence filtering condition is satisfied.

\subsection{\label{sec:level2}Coherence Imaging}
We consider spatially and temporally stationary, partially coherent optical fields to demonstrate noise-robust coherence imaging. Spatial coherence is quantified by the cross-spectral density function, $W(\boldsymbol{\rho_1}, \boldsymbol{\rho_2};\omega) = \langle E^*(\boldsymbol{\rho_1};\omega) E(\boldsymbol{\rho_2};\omega) \rangle$, where $E(\boldsymbol{\rho_1};\omega)$ and $E(\boldsymbol{\rho_2};\omega)$ represent the electric fields with frequency $\omega$ at two spatial locations, $\boldsymbol{\rho_1}$ and $\boldsymbol{\rho_2}$, respectively. This function contains all the necessary information to reconstruct the image of the object. Spatial stationarity implies that the cross-spectral density depends only on the relative displacement between the two spatial locations, $\boldsymbol{\rho_1} - \boldsymbol{\rho_2}$, and remains invariant during propagation. The different frequency components are uncorrelated as the field is also temporally stationary.

The schematic diagram of the coherence imaging setup is shown in Fig. \ref{coherence_imaging_setup}. Spatially incoherent and temporally stationary light illuminates a planar transparent object placed at the front focal plane of a converging lens. This setup generates a partially coherent, spatially and temporally stationary optical field, as depicted in Fig. \ref{coherence_imaging_setup}(a). The cross-spectral density function of the optical field at any propagation distance $z$ from the lens is given by:
\begin{equation}
\begin{aligned}
    W(\bm{\Delta\rho},z) &= \int_{-\infty}^{\infty}\int_{-\infty}^{\infty} I \left( \bm{q},\omega\right) e^{-i \bm{q} \cdot \Delta\bm{\rho}} d\bm{q} d\omega \\ & = \int_{-\infty}^{\infty}\int_{-\infty}^{\infty} I_s \left(\frac{fc}{\omega} \bm{q}\right)S(\omega) e^{-i \bm{q} \cdot \Delta\bm{\rho}} d\bm{q} d\omega,
\end{aligned}\label{csd_source_1}
\end{equation}
where $\omega$ is the frequency, $S(\omega)$ is the normalized frequency distribution of the field, $I_s \left(\frac{fc}{\omega} \bm{q}\right)$ represents the spectral density of the field which has the same functional form as the object's intensity profile, c is the speed of light, $\bm{\Delta\rho} = \bm{\rho_1} - \bm{\rho_2}$ is the spatial separation, $f$ is the focal length of the lens, and $\bm{q}$ is the transverse wave vector of the optical field (for details, see Methods section A). 

The cross-spectral density function encodes the intensity profile of the object as its spatial Fourier transform. Additionally, the partial spatial coherence of the field offers several advantages, such as speckle-free imaging \cite{redding2012np,redding2015pnas}, improved imaging through scattering media \cite{dijk2010prl,bhattacharjee2020pra}, and enhanced imaging in turbulent environments \cite{bhattacharjee2020ol}, which make it well-suited for a variety of challenging imaging and microscopy scenarios.

The image of the object can be reconstructed from the cross-spectral density function by taking its inverse spatial Fourier transform. To measure the complex-valued cross-spectral density function, we employ a phase-shifting wavefront inversion interferometer \cite{mohta2025jop}, a reference-free method that relies solely on intensity measurements to directly extract both the real and imaginary parts of the complex cross-spectral density function (for details, see Methods section B and C). The schematic of this detector is shown in Fig. \ref{coherence_imaging_setup}(c). The cross-correlation function of the incoming object field measured using this detector under quasi-monochromatic approximation (Methods section C) is,
\begin{equation}
\begin{aligned}
    \Gamma(\bm{\Delta\rho};\tau) & = \int_{-\infty}^{\infty} I_s \left(\frac{fc}{\omega_0} \bm{q}\right)e^{-i \bm{q} \cdot \Delta\bm{\rho}} d\bm{q} \int_{-\infty}^{\infty}S(\omega)e^{-i\omega\tau} d\omega\\
    & = \mathcal{F}_q \left\{I_s \left(\frac{fc}{\omega_0} \bm{q}\right)\right\} \mathcal{F}_\omega\{S(\omega)\}\\
    & = W(\Delta\bm{\rho},\omega_0)\Omega(\tau),
\end{aligned}
\label{quasi_ccf_measured_1}
\end{equation}
where $\tau$ is the time difference between the two arms of the interferometer and $\omega_0$ is the central frequency of the frequency distribution $S(\omega)$. The cross-correlation function is the product of the time-frequency Fourier transform, $\mathcal{F}_\omega$, and the space-frequency Fourier transform, $\mathcal{F}_q$.
Taking the spatial inverse Fourier transform of Eq.~\ref{quasi_ccf_measured_1} gives the image of the object.
\begin{equation}
    \mathcal{F}_{\Delta\bm{\rho}}^{-1}\{ \Gamma(\bm{\Delta\rho};\tau) \} \propto I_s \left(\frac{fc}{2\omega_0} \bm{q}\right)\Omega(\tau)
    \label{csd_inverse_ft_1}
\end{equation}

\subsection{\label{sec:level2}Principle: Noise-resilient coherence imaging}

While coherence imaging offers an alternative approach to capturing images, its significance lies in its ability to leverage coherence properties for noise rejection. The ability to image objects by measuring the spatial coherence function provides a tool to separate objects based on their coherence characteristics. This can be used to distil a clean image of an object from a noisy signal. Our technique considers two optical fields: a spectrally narrow-band field containing the object's information and another, a broader noise field. Noise can be a constant background signal or another unwanted object. 
So the spectral density of the incoming field is,
\begin{equation}
    \begin{aligned}
        I(\bm{q},\omega) &= I_{o}(\bm{q},\omega)+I_{n}(\bm{q},\omega)\\
    &=I_{s_{o}} \left(\frac{fc}{\omega} \bm{q}\right)S_{o}(\omega)+I_{s_{n}} \left(\frac{fc}{\omega} \bm{q}\right)S_{n}(\omega)
    \end{aligned}
\end{equation}
where the subscript $o$ and $n$ denote the object and noise fields. We now consider the frequency distribution $S(\omega)$ to be a normalized Gaussian function given as
\begin{equation}
    S(\omega) = \frac{2}{\sqrt{2\pi}\Delta\omega}\exp\left[-\dfrac{(\omega-\omega_0)^2}{2(\Delta\omega/2)^2}\right]
    \label{spectral_distribution}
\end{equation}
where  $\omega_0$ is the central frequency of the field and $\Delta \omega$ is the spectral bandwidth of the field.  Now, the measured cross-correlation function in Eq.~\ref{quasi_ccf_measured_1} can be expressed as,
\begin{equation}
    \begin{aligned}
        \Gamma(\bm{\Delta\rho},\tau) &= \Gamma_{o}(\bm{\Delta\rho},\tau)+\Gamma_{n}(\bm{\Delta\rho},\tau)\\
    & = W_{o}(\Delta \bm{\rho},\omega^o_0)\exp\left[-\dfrac{1}{2}{\left(\dfrac{\tau}{\tau_c^o}\right)}^2\right]\\& \qquad+ W_{n}(\Delta \bm{\rho},\omega^{n}_0)\exp\left[-\dfrac{1}{2}{\left(\dfrac{\tau}{\tau_c^n}\right)}^2\right]\\
   & = W_{o}(\Delta \bm{\rho},\omega^o_0) \exp\left[-\dfrac{1}{2}{\left(\dfrac{\Delta z}{l_c^o}\right)}^2\right]\\& \qquad+ W_{n}(\Delta \bm{\rho},\omega^{n}_0) \exp\left[-\dfrac{1}{2}{\left(\dfrac{\Delta z}{l_c^n}\right)}^2\right]
    \end{aligned}
    \label{measured_toy_ccf}
\end{equation}
where $\tau_c=2/\Delta\omega$ is the coherence time of the field, $\Delta z = c\tau$ is the path length difference between the interferometer arms and $l_c = c\tau_c$ is the temporal coherence length of the optical field. 

The temporal coherence of the field comes in as a constant multiplicative factor to the cross-spectral density function, and the temporal coherence length is inversely related to the spectral width of the field. So, the broader the spectral width, the smaller the temporal coherence length of the field and vice versa. Our distillation technique exploits the temporal coherence of the optical fields to suppress the noise field and filter out the object. To obtain high-visibility interference fringes, the path length difference between the interferometer arms must be smaller than the temporal coherence length of the optical field.  We adjust the interferometer's path length difference such that it is much shorter than the temporal coherence length of the narrow-band field but longer than that of the broad-band noise. 
\begin{equation}
    {l_c^{n}}<\Delta z < {l_c^{o}}
    \label{path_diff_condition}
\end{equation}

This suppresses the interference fringes from noise, and the measured complex cross-spectral density function contains only the spatial coherence function of the object. We set the path length difference between the arms to be $m$ times the temporal coherence length of the noise field, $\Delta z = ml_c^{n}$. Substituting into Eq.~\ref{path_diff_condition} yields the constraint $1<m<\Delta \omega_{n}/\Delta \omega_{o}$. Then, taking the spatial inverse Fourier transform, we get the filtered image of the object,
\begin{equation}
\begin{aligned}
    \mathcal{F}_{\Delta\bm{\rho}}^{-1}\{ \Gamma(\bm{\Delta\rho};\tau) \} &= I_{s_{o}} \left(\frac{fc}{\omega_0^{o}} \bm{q}\right)\exp\left[-\dfrac{m^2}{2}{\left(\dfrac{l_c^n}{l_c^o}\right)}^2\right] \\ &\qquad+ I_{s_{n}} \left(\frac{fc}{\omega_0^{n}} \bm{q}\right)\exp\left[-\dfrac{m^2}{2}\right]\\
    &= I_{s_{o}} \left(\frac{fc}{\omega_0^{o}} \bm{q}\right)\exp\left[-\dfrac{m^2}{2}{\left(\dfrac{\Delta \omega_o}{\Delta \omega_n}\right)}^2\right]\\ &\qquad+ I_{s_{n}} \left(\frac{fc}{\omega_0^{n}} \bm{q}\right)\exp\left[-\dfrac{m^2}{2}\right]
\end{aligned}
     \label{filtered_image}
\end{equation}
While the intensity of both the object and noise images is attenuated by Gaussian factors, the noise is suppressed significantly more as $\Delta \omega_{o}/\Delta \omega_{n}<1$.

\begin{figure*}[!t]
\centering
\includegraphics[scale=0.32]{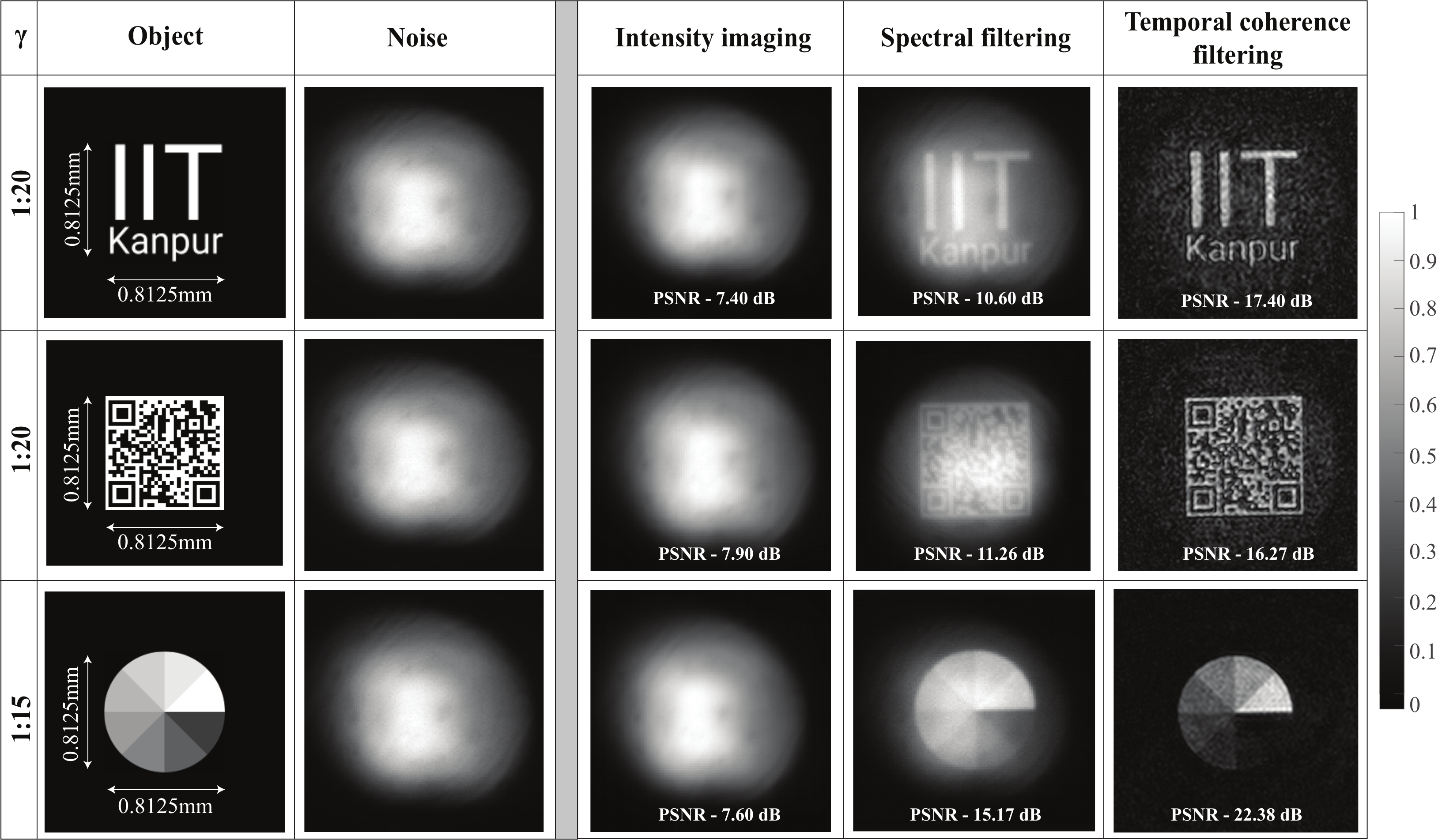}
\caption{Noise-resilient imaging with uniform noise - Three objects were imaged: Object 1: Binary plaintext image - IIT Kanpur; Object 2: QR code - a binary image with intricate structures; Object 3: Grayscale wheel with eight distinct gray levels. The parameter $\gamma$ represents the ratio of the object intensity to noise intensity. The object is illuminated with incoherent light centred at 633 nm with a 10 nm bandwidth, while the noise is illuminated with incoherent light centred at 595 nm with a 90 nm bandwidth. The size of the object is 0.8125 mm x 0.8125 mm. When imaged using a conventional lens-based system, the object is completely obscured by noise, as shown in the Intensity Imaging column. Spectral filtering, using an optical filter centred at 633 nm with a 10 nm bandwidth at the camera, rejects some noise, making the object visible. Image quality is quantified using the peak signal-to-noise ratio (PSNR). Temporal coherence filtering significantly outperforms spectral filtering, achieving superior contrast and object clarity even under extreme noise conditions. The filtered QR code can be scanned accurately, validating the method’s ability to distil complex, real-world objects from noisy environments.}
\label{unstructured_noise}
\end{figure*}
\begin{figure*}[!t]
\centering
\includegraphics[scale=0.32]{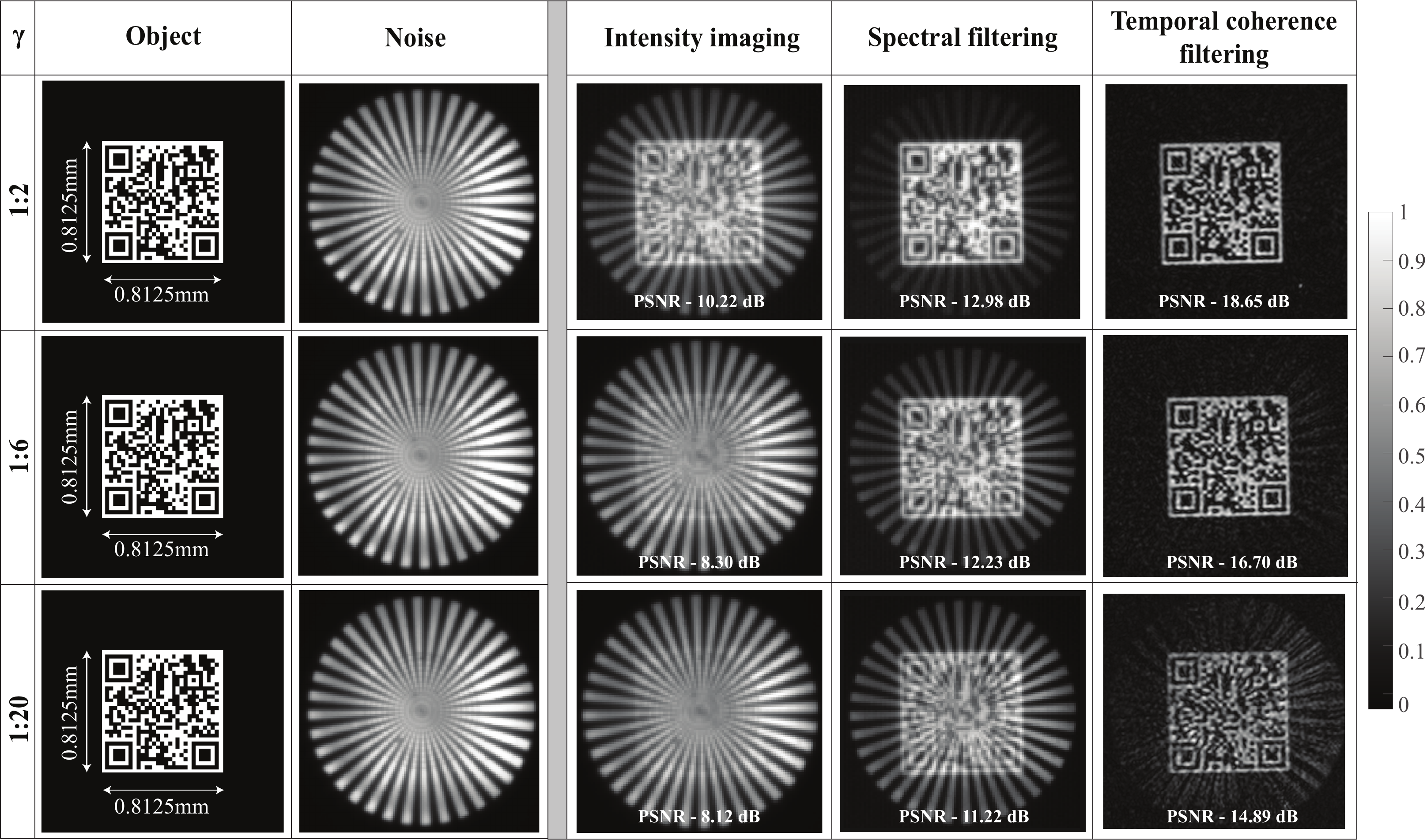}
\caption{ Noise-resilient imaging with spatially structured noise - The parameter $\gamma$ represents the ratio of the object intensity to noise intensity. The object was imaged at various noise intensities to assess the impact of noise on the distillation performance of the protocol. The object is illuminated with incoherent light centred at 633 nm with a 10 nm bandwidth, while the noise is illuminated with incoherent light centred at 595 nm with a 90 nm bandwidth. The size of the object is 0.8125 mm x 0.8125 mm. When imaged using a conventional lens-based system, the object is completely obscured by noise, as shown in the Intensity Imaging column. Spectral filtering, using an optical filter centred at 633 nm with a 10 nm bandwidth at the camera, rejects some noise, making the object visible. Image quality is quantified using the peak signal-to-noise ratio (PSNR). Temporal coherence filtering significantly outperforms spectral filtering, achieving superior contrast and object clarity even under extreme noise conditions (20× noise-to-signal ratio). The filtered QR code can be scanned accurately, validating the method’s ability to distil complex, real-world objects from noisy environments.}
\label{structured_noise}
\end{figure*}

\subsection{\label{sec:level1}Experiment: Noise-resilient coherence imaging}
For our experiment, we generate the spatially and temporally stationary partially coherent field starting with a light-emitting diode (LED), as LEDs can be considered spatially incoherent and temporally stationary source~\cite{tziraki2000apb,diLorenzoPires2010ol,aarav2017pra,torcalMilla2022optik,deng2017scr}. An OLED display of a smartphone serves as the incoherent source, and the displayed image acts as the object. Equivalently, an LED imaged onto a transparent object can also act as the object. It is positioned at the front focal plane of a converging lens with a focal length of $f=500\text{mm}$, and the outgoing optical field is filtered through an optical filter with a $10$nm spectral width centred at $633$nm, resulting in an effective temporal coherence length of approximately $35\mu \text{m}$. It is mixed with noise at the beam-splitter BS1, as shown in Fig. \ref{coherence_imaging_setup}(b), resulting in the superposition of the two images. The noise is generated using an LED source with a spectral width of $90$nm, centred at $595$nm, illuminating a transparent amplitude object and placed at the front focal plane of a converging lens with focal length $f=500\text{mm}$. It has a temporal coherence length of approximately $3.5\mu \text{m}$. 
The average photon flux of the object field at the camera was approximately 1.7 photons per frame per pixel. Due to the low photon flux, an EMCCD camera was used with an exposure time of 50 ms and EM gain of 300. The data was averaged over 5,000 frames to achieve the signal-to-noise ratio required for reliable reconstruction, resulting in an acquisition time of about 4 minutes per phase step. The acquisition time is limited by the source brightness and can be reduced by increasing the illumination or using a higher-efficiency detector.

The path length difference between the interferometer arms is adjusted using the two mirrors mounted on the translation stage (TS), allowing micron-level adjustments. The path-length difference is tuned to be shorter than the temporal coherence length of the narrowband object field but longer than that of the broadband noise.
To determine and optimize the path-length difference $\Delta z$, we first use a square object whose cross-spectral density is analytically known to be a two-dimensional sinc function. One arm of the interferometer is translated using a translation stage while the geometric phase is adjusted to maximize the output intensity at the camera. For each translation stage position, the corresponding fringe pattern is recorded and its visibility computed. The position yielding maximum fringe visibility corresponds to the zero path-length difference ($\Delta z=0$) within the precision of the translation stage. Once this reference is established, the stage can be moved by controlled increments to set any desired $\Delta z$ for temporal-coherence filtering. In our experiment, the translation stage has a positioning precision of 0.5 µm. The optimization is performed once before measurements, and because the interferometer is vibration-isolated and enclosed, the optimized $\Delta z$ remains stable over the entire acquisition period.

Filtering out noise in this manner while simultaneously measuring the cross-spectral density function is possible due to the unique controls in our interferometer. As detailed in Sections B and C of the Methods, measuring the complex cross-spectral density function requires recording interferograms at multiple fixed phase differences between the interfering fields. The implementation of the temporal coherence filtration technique is made possible because the phase difference is controlled not by changing the path length difference of the interferometer—as is the standard practice—but by introducing a geometric phase difference through a combination of a polarizer, two quarter-wave plates, and a half-wave plate, Fig.~\ref{coherence_imaging_setup}(c). The use of geometric phase enables the path length difference between the interferometer arms to be utilised for noise filtering and also provides more precise and easier control of the phase difference.

Experiments have been performed with both spatially uniform and structured noise. The experimental results with uniform noise are shown in Fig. \ref{unstructured_noise}. Three different objects were imaged: (1) A binary plaintext image of IIT Kanpur, (2) a QR code - a binary object with intricate structures, and (3) a grayscale wheel with eight grayscale levels. The objects were approximately 0.8125 mm x 0.8125 mm in size. We further tested our technique with spatially structured noise. The object used was a QR code, and it was imaged at various noise intensities to assess the effect of noise on the distillation performance of the filtering technique. The experimental results are shown in Fig. \ref{structured_noise}. The first column is relative noise intensity, which is quantified by the signal-to-noise ratio (SNR) of the unfiltered field, $\gamma = \text{object intensity}:\text{noise intensity}$. The second and third columns contain the object and the noise, respectively. The fourth column contains images acquired using a standard lens-based imaging system. The fifth column shows images obtained with spectral filtering, using a $10$nm optical filter centred at $633$nm placed at the camera. The final column displays the images acquired using coherence imaging, employing the temporal coherence filtering protocol. To quantitatively assess reconstruction quality, we employ the peak signal-to-noise ratio (PSNR). The measured spatial coherence functions, used to reconstruct the temporal coherence-filtered images in Figs. \ref{unstructured_noise} and \ref{structured_noise}, are provided in the supplemental document \cite{supplemental}. The results demonstrate that image distillation through temporal coherence filtering successfully recovers both intricate structures and varying shades under noise with intensity as large as 20 times that of the signal. It also significantly outperforms the spectral filtering technique. The filtered QR code can be scanned accurately, validating the method’s ability to distil complex, real-world objects from noisy environments.

\subsection{\label{sec:level1}Performance comparison: temporal coherence filtering versus spectral filtering}

Since the bandwidth of the light from the object is smaller than that of the noise, $\Delta \omega_{o} < \Delta \omega_{n}$, spectral filtering using a bandpass optical filter is a potent and straightforward way of reducing noise. It is, therefore, essential to compare the performance of the two filtering techniques. We consider the situation where the object and noise are the same binary image with the central frequencies $\omega^o_0$ and $\omega^n_0$. When no filtering technique is applied, the acquired image is,
\begin{equation}
    I_\text{image}(\bm{\rho})=\frac{A_{o}}{\sigma}J_{o}(\bm{\rho}) + \frac{A_{n}}{\sigma}J_{n}(\bm{\rho})
\end{equation}
where $J(\bm{\rho)}$ is the normalized spatial intensity distribution, $A$ is the power of the field and $\sigma$ is the area of the image. The higher the SNR ratio $\gamma=A_o/A_{n}$, the lower the noise in the acquired image and the better the distillation.

For spectral filtering (SF), consider an optical bandpass filter centred at $\omega_0$ having a Gaussian profile with bandwidth $\Delta \omega_o$. Such a filter profile is considered because the frequency distribution of the object and noise field was considered to be normalized Gaussian distributions (Eq.~\ref{spectral_distribution}). The image acquired using this filtering is,
\begin{equation}
\begin{aligned}
        I_\text{image}(\bm{\rho})=\frac{A_o}{\sqrt{2}\sigma}J_{s_{o}}(\bm{\rho}) + \frac{A_{n}}{\sigma}\frac{\Delta \omega_{o}}{\Delta \omega_{n}}\exp\left[\dfrac{-2(\omega^o_0-\omega^n_0)^2}{\Delta\omega^2_o+\Delta\omega^2_n}\right]\cdot\\\cdot\left[1+\left(\frac{\Delta \omega_o}{\Delta \omega_n}\right)^2\right]^{-\frac{1}{2}}J_{s_{n}}(\bm{\rho})
\end{aligned}
\end{equation}
The performance of spectral filtering is,
\begin{equation}
    \gamma_\text{sf}=\frac{A_{o}}{\sqrt{2}A_{n}}\frac{\Delta \omega_{n}}{\Delta \omega_{o}}\exp\left[\dfrac{2(\omega^o_0-\omega^n_0)^2}{\Delta\omega^2_o+\Delta\omega^2_n}\right]\left[1+\left(\frac{\Delta \omega_o}{\Delta \omega_n}\right)^2\right]^{\frac{1}{2}} \label{performance_sf}
\end{equation}

\begin{figure*}[!t]
\centering
\includegraphics[scale=1.2]{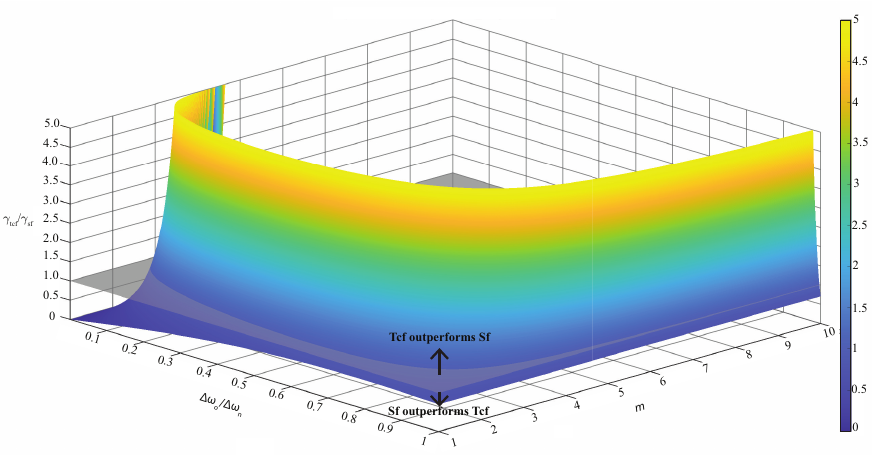}
\caption{Comparison of Temporal Coherence Filtering (TCF) and Spectral Filtering (SF) for the same-central-frequency case – Plot of the ratio $\gamma_\text{tcf}/\gamma_\text{sf}$ as a function of $m$ and $\Delta \omega_{o}/\Delta \omega_{n}$. The ratio $\gamma_\text{tcf}/\gamma_\text{sf}$ quantifies the relative effectiveness of TCF over SF in suppressing noise, where $m$ denotes the factor by which the path-length difference between the interferometer arms exceeds the temporal-coherence length of the noise field. A ratio greater than 1 indicates that TCF outperforms SF. The plot demonstrates that there exists a threshold value of $m$ for each $\Delta \omega_o/\Delta \omega_{n}$ beyond which TCF outperforms SF.}
\label{tcf_vs_sf}
\end{figure*}

Next, consider temporal coherence filtering (TCF) with the path length difference between the arms to be $m$ times the temporal coherence length of the noise field, $\Delta z = ml_c^{\text{\scalebox{1}{\textit{n}}}}$. The acquired image follows from Eq.~\ref{filtered_image},
\begin{equation}
\begin{aligned}
    I_\text{image}(\bm{\rho})=\frac{A_o}{\sigma}J_{s_o}(\bm{\rho})\exp\left[-\dfrac{m^2}{2}{\left(\dfrac{\Delta \omega_o}{\Delta \omega_n}\right)}^2\right]\\ + \frac{A_n}{\sigma}J_{s_n}(\bm{\rho})\exp\left[-\dfrac{m^2}{2}\right]
\end{aligned} 
\end{equation}
The performance of temporal coherence filtering is,
\begin{equation}
    \gamma_\text{tcf} = \dfrac{A_{o}}{A_{n}}\exp\left[\dfrac{m^2}{2}\left\{1-\left(\dfrac{\Delta \omega_{{o}}}{\Delta \omega_{n}}\right)^2\right\}\right]
\end{equation}
To compare the performance of the two filtering techniques, we take the ratio of their SNRs,
\begin{equation}
    \begin{split}
        \frac{\gamma_\text{tcf}}{\gamma_\text{sf}}=&\dfrac{\sqrt{2}\Delta \omega_{o}}{\Delta \omega_{n}}\exp\left[\dfrac{m^2}{2}\left\{1-\left(\dfrac{\Delta \omega_{{o}}}{\Delta \omega_{n}}\right)^2\right\}\right]\\&\exp\left[\dfrac{-2(\omega^o_0-\omega^n_0)^2}{\Delta\omega^2_n\{1+(\Delta\omega_o/\Delta\omega_n)^2\}}\right]\left[1+\left(\frac{\Delta \omega_o}{\Delta \omega_n}\right)^2\right]^{-\frac{1}{2}}
    \end{split}
    \label{filter_ratio}
\end{equation}
When the above ratio is greater than 1, temporal coherence filtering performs better than spectral filtering and vice versa. Eq.~\ref{filter_ratio} shows that in general, the SNR ratio between spectral filtering and temporal-coherence filtering depends on both the bandwidth ratio and the absolute central wavelengths of the signal and noise fields. For given central frequencies and bandwidths, the SNR ratio increases monotonically with the parameter $m = \Delta z/l^n_c$, reflecting the fact that larger path-length differences enhance the relative suppression of the noise field in temporal-coherence filtering. When the two fields have the same central frequency, the SNR ratio depends solely on the bandwidth ratio. This equal-central-frequency condition provides a baseline for comparing the intrinsic performance of temporal coherence and spectral filtering. Accordingly, Fig.~\ref{tcf_vs_sf} shows the plot of $\gamma_{\text{tcf}}/\gamma_{\text{sf}}$ as a function of $m$ and $\Delta \omega_{o}/\Delta \omega_{n}$ for this case. The plot demonstrates that there exists a threshold value of $m$ for each $\Delta \omega_o/\Delta \omega_{n}$ beyond which temporal coherence filtering outperforms spectral filtering.

From Eqs.~\ref{performance_sf} and \ref{filter_ratio}, it follows that when the central frequencies of the object and noise fields are well separated and the spectral overlap is small, spectral filtering is the preferred approach. However, when the central frequencies are close and the spectra substantially overlap, the effectiveness of spectral filtering diminishes. In contrast, temporal-coherence filtering—whose performance is independent of the central frequencies and of the spectral overlap, as shown by Eq.~\ref{filter_ratio}—remains effective in this regime and outperforms spectral filtering.

\section{\label{sec:level5}Conclusion}

We present a coherence filtering technique that effectively separates object from noise based on their temporal coherence properties. We implement this using our interferometric measurement protocol, which reconstructs images through spatial coherence measurement while simultaneously filtering out noise using temporal coherence. This capability enables noise-resilient imaging, even in high-noise environments. We experimentally demonstrate the effectiveness of our approach using objects with rich features (QR code) and eight gray levels (grayscale wheel), subjected to spatially uniform and structured noise at intensities up to 20 times that of the object. The 20-fold noise intensity is not an upper limit; the method’s noise-handling capability depends on the difference in temporal coherence lengths between the object and noise fields, which in turn depends on their respective frequency bandwidths. With an average photon flux of 1.7 photons/frame/pixel of the object field at the camera, our system operates efficiently under low-light conditions. Thus making our method highly suitable for extreme low-light imaging scenarios.

Our method overcomes key limitations commonly faced by traditional techniques, including reliance on prior knowledge of noise characteristics, difficulties with spatially structured noise, and challenges in balancing noise reduction with signal preservation. Spectral filtering offers good noise suppression when the object and noise spectra are well separated, but its utility diminishes when their spectra overlap. In this regime, temporal-coherence filtering provides a robust alternative, maintaining strong noise rejection where spectral filtering fails. Our technique seamlessly integrates into standard imaging setups, overcoming the need for extensive modifications or specialized illumination, unlike quantum-based approaches. Moreover, whereas quantum techniques typically require hours of data acquisition, our method achieves reliable reconstructions within minutes, offering a substantially more practical and time-efficient solution.

Building on these advantages, our method extends beyond objects emitting light at a single central wavelength (i.e., a single colour band). The theoretical model suggests that the method can also be applied to multicoloured objects emitting light at multiple central wavelengths simultaneously, provided each wavelength has a narrow spectral bandwidth compared to noise, with minimal overlap between bands. This allows for imaging of multicoloured objects in a single measurement, without the need for separate measurements for each colour band. This represents a significant improvement over quantum distillation techniques, which are not applicable to multicoloured objects, and optical filter-based spectral filtering, which requires separate measurements for each colour band. The ability to image multicoloured objects in a single measurement is crucial for applications such as biological imaging and fluorescence microscopy, where multicoloured labels are commonly used.

By rejecting noise without requiring prior knowledge of object or noise characteristics and its ability to function at low-light levels, temporal coherence filtering holds great potential for advancing a wide range of applications, including fluorescence microscopy, biological imaging, and optical communication, positioning it as a valuable tool for both research and practical uses across diverse fields.

\section{\label{sec:level1}Methods}

\subsection{\label{sec:level1}Object illumination}

We demonstrate the coherence imaging technique using a propagation-invariant, spatially and temporally stationary, partially coherent field. Consider a monochromatic, planar, spatially incoherent primary source, positioned at the front focal plane ($z = -f$) of a lens (L2) located at $z = 0$, Fig.~\ref {coherence_imaging_setup}(a). The combined system of the planar source and lens forms our spatially partially coherent field source. The field originating from any spatial point $\bm{\rho}'$ at a distance $z$ is denoted as $E_s(\bm{\rho}', z)$.

Due to the spatial incoherence of the primary source, the fields emerging from two distinct spatial points, $\bm{\rho}'_1$ and $\bm{\rho}'_2$, are completely uncorrelated. This relationship is described by:
\begin{equation}
\langle E_s^*(\bm{\rho}'_1, -f) E_s(\bm{\rho}'_2, -f) \rangle_e = I_s(\bm{\rho}'_1, -f)\delta(\bm{\rho}'_1 - \bm{\rho}'_2)\label{spatial_uncorrelation}
\end{equation}
where $I_s(\bm{\rho}'_1, -f)$ is the intensity at point $\bm{\rho}'_1$.

For our incoherent source, we employ a smartphone with OLED screen, as light-emitting diodes (LEDs) can be considered spatially completely incoherent in the sense that their position correlations are approximated by the form given in Eq.~\ref{spatial_uncorrelation} ~\cite{tziraki2000apb,diLorenzoPires2010ol,aarav2017pra,torcalMilla2022optik,deng2017scr}. Positioned at the front focal plane of a convex lens, these independent point sources transform into plane waves with corresponding amplitudes, $a(\bm{q_1})$, where $\bm{q_1}$ represents the transverse wave vector. The lens converts the spatial incoherence into angular incoherence, represented by the angular correlation function $\mathcal{A}(\bm{q_1}, \bm{q_2})$: 
\begin{equation}
\mathcal{A}(\bm{q_1}, \bm{q_2}) = \langle a^*(\bm{q_1}) a(\bm{q_2}) \rangle_e = I(\bm{q_1}) \delta(\bm{q_1} - \bm{q_2})
\end{equation}
Here, $I(\bm{q_1})$ represents the spectral density of the field, exhibiting the same functional form as the source intensity. This angular correlation is crucial to the spatial stationarity and propagation invariance of the generated partially coherent field.

Following the approach in~\cite{mandel1995coherence}, the cross-spectral density function $W(\bm\rho_1, \bm\rho_2, z)\equiv \langle E^*(\bm\rho_1, z) E(\bm\rho_2, z) \rangle_e$ at $z=z$ is 
\begin{align} W({\bm{\rho}_1}, {\bm{\rho}_2}, z)= & \iint_{-\infty}^{\infty} \mathcal{A}(\bm{q_1}, \bm{q_2}) \notag\\ \times & e^{-i\bm{q_1}.\bm\rho_1+i\bm{q_2}.\bm\rho_2} e^{-i\tfrac{(q_1^2-q_2^2)z}{2k_0}}d\bm{q_1} d\bm{q_2}\label{cs-density} 
\end{align}
Eq.~\ref{cs-density} dictates the evolution of spatial correlations in the field, as denoted by the cross-spectral density function, during propagation in the region $z>0$ after passing through the lens.
By substituting the angular correlation function into this equation, we obtain the cross-spectral density:
\begin{equation}
W(\bm{\Delta\rho}, z) = \int_{-\infty}^{\infty} I(\bm{q}) e^{-i\bm{q} \cdot \bm{\Delta\rho}} d\bm{q}\label{cs-density2}
\end{equation}
where $\bm{\Delta\rho} = \bm{\rho}_1 - \bm{\rho}_2$.

The previous discussion applied to a monochromatic field. For a broadband field, assuming the optical field is temporally stationary (i.e., its frequency components are uncorrelated), the cross-spectral density function is:
\begin{equation}
\begin{aligned}
    W(\bm{\Delta\rho},z) &= \int_{-\infty}^{\infty}\int_{-\infty}^{\infty} I \left( \bm{q},\omega\right) e^{-i \bm{q} \cdot \Delta\bm{\rho}} d\bm{q} d\omega \\ & = \int_{-\infty}^{\infty}\int_{-\infty}^{\infty} I_s \left(\frac{fc}{\omega} \bm{q}\right)S(\omega) e^{-i \bm{q} \cdot \Delta\bm{\rho}} d\bm{q} d\omega,
\end{aligned}\label{cc_func_source}
\end{equation}
where $\omega$ is the frequency, $S(\omega)$ is the normalized frequency distribution of the field, $I_s \left(\frac{fc}{\omega} \bm{q}\right)$ represents the spectral density of the field which exhibits the same functional form as the source intensity and c is the speed of light. The corresponding intensity of the field is then given by:
\begin{equation}
I(\bm{\rho}, z) = W(\bm{\rho}, \bm{\rho}, z)=\int_{-\infty}^{\infty} \int_{-\infty}^{\infty}I(\bm{q}, \omega) d\bm{q}d\omega=K
\end{equation}
The cross-spectral density function \( W(\bm{\Delta\rho}, z) \) in Eq.~(\ref{cs-density2}) is expressed in terms of coherent mode representation, where the coherent modes correspond to plane waves. The resulting field demonstrates three key features: (1) \textit{Propagation invariance}—the cross-spectral density function and intensity remain constant regardless of changes in \(z\); (2) \textit{Spatial stationarity}—the intensity \(I(\bm{\rho}, z)\) is independent of position \(\bm{\rho}\), while the cross-spectral density function depends only on the separation \(\bm{\Delta\rho}\); and (3) \textit{Fourier relationship}—the cross-spectral density function \(W(\bm{\Delta\rho}, z)\) is the Fourier transform of the spectral density \(I(\bm{q})\). Since the spectral density reflects the intensity distribution of the primary source, this implies that the cross-spectral density function is essentially the Fourier transform of the primary source’s intensity profile. Notably, this imposes no restrictions on the form of the intensity function \(I_s(\bm{\rho}')\) that the incoherent source can exhibit.

\subsection{\label{sec:level1}Coherence function measurement}

To measure the complex cross-spectral density function, we employ our phase-shifting wavefront-inversion interferometry technique. In this method, the optical field and its inverted counterpart are superimposed, and interferograms are recorded at four distinct phase differences between the arms of the interferometer. This provides both the real and imaginary components of the cross-spectral density function.

As depicted in Fig.~\ref{coherence_imaging_setup}(c), the setup employs a Mach-Zehnder type interferometer. The key components of this system are the two Dove prisms that invert the wavefront of the incoming optical field. Dove prisms can cause corner obstructions and front surface reflections. To address these problems, K-mirrors, as described in \cite{karan2022ao}, can be used as an alternative. The geometric phase unit, consisting of two quarter-wave plates oriented at \(45^\circ\) with respect to the polarizer at the entrance of the interferometer, and a half-wave plate (HWP) placed between them, in one of the arms, is used for modulating the phase difference between the arms. Rotating the HWP through an angle \(\theta\), introduces a geometric phase shift of \(2\theta\) \cite{jha2008prl}. The resulting electric field at the interferometer's output port is
\begin{multline}
    V(x,y;t) = {k_1}V(-x,y;t-t_1) + {k_2}V(x,-y;t-t_2)\\=k_1\int_{-\infty}^{\infty}E(-x,y;\omega)e^{-i[\omega(t-t_1)+\beta_1]}d\omega \\+ k_2\int_{-\infty}^{\infty}E(x,-y;\omega)e^{-i[\omega(t-t_2)+\beta_2]}d\omega
\end{multline}
where \(t_1\) and \(t_2\) represent the travel times of the optical field through the two arms, \(\beta_1\) and \(\beta_2\) are additional phase terms, and \(k_1\) and \(k_2\) are constants. For temporally stationary fields,
\small
\begin{multline}
    \langle E^*(-x,y;\omega')E(x,-y;\omega)\rangle_e=W(-x,y;x,-y;\omega)\delta(\omega-\omega')
\end{multline}
\normalsize
where $W(-x,y;x,-y;\omega)$ is the cross-spectral density function. The intensity at the camera plane is,
\small
\begin{align}
I_{\rm out}(x, y;\tau) = & \ k_1^2I_1(-x,y)+k_2^2I_2(x,-y) \nonumber \\
& + 2k_1 k_2 \text{Re}[\Gamma(-x,y;x,-y;\tau)] \cos \delta \nonumber \\
& - 2k_1 k_2 \text{Im}[\Gamma(-x,y;x,-y;\tau)] \sin \delta, \label{intensity_out}
\end{align}
\normalsize
where $\tau=t_1-t_2$, $\delta = \beta_2-\beta_1$,\\ $I_i(x,y;t-t_i)=\int_ {-\infty}^{\infty}|E_{in}(x,y;\omega)|^2 d\omega$ for $i=1,2$ are the intensities from the two arms, \\ $\Gamma(-x,y;x,-y;\tau)=\int_ {-\infty}^{\infty} W(-x,y;x,-y;\omega)e^{-i\omega\tau}d\omega$ denotes the cross-correlation function, which is the Fourier transform of the cross-spectral density function. The \(\text{Re}[\cdots]\) and \(\text{Im}[\cdots]\) represents the real and imaginary parts, respectively.

To isolate the cross-correlation function, the output intensities are measured at four specific values of \(\delta\). For phase differences \(\delta = \delta_c\) and \(\delta = \delta_d\), the difference intensity \(\Delta I_{\rm out}(x, y)\) is:
\begin{multline}
\Delta I^{(\delta_c, \delta_d)}_{\rm out}(x, y;\tau)= 2k_1k_2 \lbrace {\rm Re}[\Gamma(-x,y;x,-y;\tau)]\\(\cos\delta_c-\cos\delta_d) 
\\-{\rm Im}[\Gamma(-x,y;x,-y;\tau)](\sin\delta_c-\sin\delta_d)\rbrace.\label{interferometer_intensity_diff}
\end{multline}
By setting \(\delta_c = 0\) and \(\delta_d = \pi\), we extract real part,
\begin{equation}
    \text{Re}[\Gamma(-x, y; x, -y;\tau)] = \frac{\Delta I_{\rm out}^{(0, \pi)}(x, y;\tau)}{4k_1 k_2}\label{ccf_real}
\end{equation}
Similarly, setting \(\delta_c = 3\pi/2\) and \(\delta_d = \pi/2\) allows us to retrieve the imaginary part,
\begin{equation}
    \text{Im}[\Gamma(-x, y; x, -y;\tau)] = \frac{\Delta I_{\rm out}^{(3\pi/2, \pi/2)}(x, y;\tau)}{4k_1 k_2}\label{ccf_imag}
\end{equation}

Thus, by recording the difference intensities for these specific phase values, we obtain both the real and imaginary components of the cross-spectral density function with a common scaling factor of \(4k_1 k_2\). This approach simplifies the measurement by eliminating the need for precise knowledge of the constants \(k_1\), \(k_2\), \(I_1\), and \(I_2\).

\subsection{\label{sec:level2}Image reconstruction through coherence function measurement}

The cross-spectral density function of the optical field encodes the intensity profile of the object via a Fourier transform relation, as seen from Eq.~\ref{cc_func_source}. The cross-correlation function of the incoming field measured using the aforementioned interferometer is
\begin{equation}
\begin{aligned}
    \Gamma(\bm{\Delta\rho};\tau) &= \int_{-\infty}^{\infty}\int_{-\infty}^{\infty} I \left( \bm{q},\omega\right) e^{-i \bm{q} \cdot \Delta\bm{\rho}} d\bm{q}e^{-i\omega\tau} d\omega \\ & = \int_{-\infty}^{\infty}\int_{-\infty}^{\infty} I_s \left(\frac{fc}{\omega} \bm{q}\right)S(\omega) e^{-i \bm{q} \cdot \Delta\bm{\rho}} d\bm{q} e^{-i\omega\tau} d\omega,
\end{aligned}\label{cc_func_measured}
\end{equation}
here $\Delta\bm{\rho} \equiv (2x,-2y)$.

There are two Fourier transforms involved in this process. One is the time-frequency Fourier transform, $\mathcal{F}_\omega$, and the other is the space-frequency Fourier transform, $\mathcal{F}_q$.
\begin{equation}
    \Gamma(\bm{\Delta\rho},\tau) = \mathcal{F}_\omega\{W(\Delta\bm{\rho},\omega)\}=\mathcal{F}_\omega\{\mathcal{F}_q\{I(\bm{q},\omega)\}\}
\end{equation}\label{two_fft}

Further, assuming the field's spectrum to be quasi-monochromatic, i.e., the spectral width $\Delta \omega$ of the source is very small compared to the central frequency $\omega_0$ of the source, $\Delta \omega\ll\omega_0$, $I_s \left(\frac{fc}{\omega} \bm{q}\right)$ can be approximated as $I_s \left(\frac{fc}{\omega_0} \bm{q}\right)$. The correlation function can be written as,
\begin{equation}
\begin{aligned}
    \Gamma(\bm{\Delta\rho};\tau) & = \int_{-\infty}^{\infty} I_s \left(\frac{fc}{\omega_0} \bm{q}\right)e^{-i \bm{q} \cdot \Delta\bm{\rho}} d\bm{q} \int_{-\infty}^{\infty}S(\omega)e^{-i\omega\tau} d\omega\\
    & = \mathcal{F}_q \left\{I_s \left(\frac{fc}{\omega_0} \bm{q}\right)\right\} \mathcal{F}_\omega\{S(\omega)\}\\
    & = W(\Delta\bm{\rho},\omega_0)\Omega(\tau),
\end{aligned}\label{quasi_cc_func_measured_2}
\end{equation}
The cross-correlation function factorizes into the product of the cross-spectral density function and the temporal correlation function. The effect of partial temporal coherence appears as a constant multiplicative factor with the cross-spectral density function.
The image can be reconstructed by measuring the cross-correlation function and performing the inverse spatial Fourier transform.

Accurately capturing the real and imaginary parts of the cross-correlation requires recording interferograms precisely at the four specific phase differences between the interferometer arms, Eq.~[\ref{ccf_real},\ref{ccf_imag}].
While precise phase determination is essential for measuring the complex cross-correlation function, it is not critical for image reconstruction. In fact, the measurement can begin at any initial phase difference between the interferometer arms. The only requirement is that subsequent interferograms are recorded at consistent phase intervals of $\pi/2$.

To see this, let us start the measurement at phase difference $\alpha$ instead of $0$. This modifies the difference intensity Eq.~\ref{interferometer_intensity_diff} to
\begin{multline}
\Delta I_\text{out}(x, y;\tau)=2k_1k_2\lbrace\text{Re}[\Gamma(-x,y;x,-y;\tau)]( \text{cos} \delta_c \text{cos} \alpha\\-\text{sin} \delta_c \text{sin} \alpha - \text{cos} \delta_d \text{cos} \alpha+ \text{sin} \delta_d \text{sin} \alpha ) 
\\-\text{Im}[\Gamma(-x,y;x,-y;\tau)]( \text{sin} \delta_c \text{cos} \alpha  \\+\text{cos} \delta_c \text{sin} \alpha - \text{sin} \delta_d \text{cos} \alpha -\text{cos} \delta_d \text{sin} \alpha )\rbrace\label{interferometer_intensity_diff_with_error}
\end{multline}
Setting $\delta_c=0$ and $\delta_d=\pi$ we get,
\begin{multline}
\Delta I_\text{out}(x, y;\tau)|_1=4k_1k_2 \lbrace (\text{cos} \alpha)\text{Re}[\Gamma(-x,y;x,-y;\tau)]\\-(\text{sin} \alpha)\text{Im}[\Gamma(-x,y;x,-y;\tau)]\rbrace.
\end{multline}
Setting $\delta_c=3\pi/2$ and $\delta_d=\pi/2$ we get,
\begin{multline}
\Delta I_\text{out}(x, y;\tau)|_2=4k_1k_2 \lbrace (\text{sin} \alpha)\text{Re}[\Gamma(-x,y;x,-y;\tau)]\\+(\text{cos} \alpha)\text{Im}[\Gamma(-x,y;x,-y;\tau)]\rbrace.
\end{multline}
The measured cross-correlation function is, 
\begin{equation}
    \begin{aligned}
        \Gamma_\text{M}(-x,y;x,-y;\tau)&=\Delta I_\text{out}(x, y;\tau)|_1 + i\Delta I_\text{out}(x, y;\tau)|_2\\
        &=4k_1k_2\Gamma(-x,y;x,-y;\tau)e^{i\alpha}
    \end{aligned}
\end{equation}
Taking the absolute value of the inverse spatial Fourier transform of the above equation yields the image of the object $I_s(\bm{\rho}')$,
\begin{equation}
   |\mathcal{F}_{\Delta\bm{\rho}}^{-1}\{ \Gamma_\text{M}(2x,-2y;\tau) \}| = 8k_1k_2 I_s \left(\frac{fc}{2\omega_0}q_x,-\frac{fc}{2\omega_0}q_y\right)|\Omega(\tau)|
\end{equation}\label{reconst_image}
This ability to reconstruct the image by measuring the cross-correlation starting at any random phase difference between the interferometer’s arms offers a significant advantage. It not only eliminates the need to precisely determine and set the phase difference to a specific value but also allows for the measurement of multicoloured objects using the same setup without additional steps.

\section{\label{sec:level1}Acknowledgement}

We thank Kalash Talati for providing the smartphone used in the experiment and Ryo Mizuta Graphics for the 3D assets used in the illustrations and experimental setup figures. We acknowledge financial support from the Science and Engineering Research Board through grants STR/2021/000035 and CRG/2022/003070, from the Department of Science \& Technology, Government of India through grant DST/ICPS/QuST/Theme-1/2019 and through the National Quantum Mission (NQM) technical group (TG) project on quantum imaging.
P.M. thanks the Prime Minister Research Fellowship (PMRF), Government of India, for financial support.

\section{\label{sec:level1}Supplemental Document}

The measured spatial coherence functions used to reconstruct the temporal coherence-filtered images, as well as analysis of the effects of path length drift and phase step errors on filtering and image reconstruction, are provided in the Supplemental document.

\bibliography{EncDec_ref}

\end{document}